\documentclass[showpacs,aps,graphicx,twocolumn]{revtex4}
\usepackage{graphicx}

\begin{document}

\title{Blind quantum computation with noise environment}

\author{Yu-Bo Sheng,$^{1}$\footnote{shengyb@njupt.edu.cn} Lan Zhou $^{2}$}
\address{$^1$Key Lab of Broadband Wireless Communication and Sensor Network
 Technology,
 Nanjing University of Posts and Telecommunications, Ministry of
 Education, Nanjing, 210003,
 China\\
 $^2$College of Mathematics \& Physics, Nanjing University of Posts and Telecommunications, Nanjing,
210003, China\\}

\date{\today }
\begin{abstract}
Blind quantum computation (BQC) is a new type of quantum computation model. BQC allows a client (Alice) who does not have enough
sophisticated technology and knowledge to perform universal quantum computation and resorts a  remote quantum computation server (Bob) to delegate universal quantum computation. During the computation, Bob cannot know Alice's inputs, algorithm and outputs. In single-server BQC protocol, it requires Alice to prepare and distribute single-photon states to Bob. Unfortunately, the distributed single photons will suffer from noise, which not only makes the single-photon state decoherence, but also makes it loss.  In this protocol, we describe an anti-noise BQC protocol, which combined the ideas of faithful distribution of single-photon state in collective noise, the feasible quantum nondemolition measurement and Broadbent-Fitzsimons-Kashefi (BFK) protocol. This protocol has several advantages. First, Alice does not require any auxiliary resources, which reduces the client's economic cost. Second, this protocol not only can protect the state from the collective noise, but also can distill the single photon from  photon loss.  Third, the noise setup in Bob is based on the linear optics, and it is also feasible in experiment. This anti-noise BQC may show that it is possible to perform the BQC protocol in a noisy environment.
\end{abstract}
 \maketitle
\section{Introduction}
Quantum computation has attracted much interest for its ultrafast computation ability. Shor's algorithm for integer factorization \cite{shor}, Grover's algorithm
 and the optimal Long's algorithm for unsorted
database search \cite{grover,long}, all have displayed the great
computing power of quantum computers. Small-scale
quantum computers in ions \cite{ion},  superconduction \cite{superconduct}, photons \cite{photon}, and some other important quantum systems have been widely investigated \cite{other1}. It is not a dream to successfully product a quantum computer in the  foreseeable future. Like current supercomputers, the first generation of quantum computers must be very expensive and  owned by very few governments or  big companies. As an ordinary quantum computer client,  he or she has poor quantum ability and are insufficient to realize universal quantum computation. Certainly, he or she can resort the quantum computation server's help to realize the computation. Moreover, he or she should protect his data without leakage.  Blind quantum computation (BQC) is a new type of quantum computation model that the client who does not have enough quantum knowledge and sophisticated technology and resorts the quantum computation servers to perform the universal
quantum computation. During the computation, the client's inputs, algorithms and outputs should be absolutely security.

In 2005, Childs proposed the first BQC model \cite{blind1}. It
is the standard quantum circuit model.  Bob needs to perform the quantum gates and Alice
requires the quantum memory. In 2009,  Broadbent, Fitzsimons, and Kashefi (BFK) proposed a BQC protocol based on the one-way quantum computation model \cite{blind2}.
 In their protocol, Alice only requires
to generate the single-qubit quantum state and a classical computer. The most advantage of this protocol is that Alice does not need the quantum memory. There are also some other important BQC protocols \cite{blind3,blind4,blind5,blind6,blind7,blind8,blind9,blind10,blind11,blind12,blind13,blind14,blind15,blind16,blind17,blind18,blind19,blind20,blind21}. For example, Morimae \emph{et al.}
proposed two BQC protocols based on the Affleck-Kennedy-Lieb-Tasaki state \cite{blind3}. Fitzsimons and Kashefi constructed
a new verifiable BQC protocol \cite{blind5}. The experiment  of the BFK protocol based
on the optical system was also reported \cite{blind10}.  Generally, these kinds of BQC protocols can be divided into three groups.   The first group is  the single-server BQC model \cite{blind1,blind2,blind3,blind5,blind6,blind7,blind8,blind9,blind10,blind12,blind13,blind14,blind15,blind16,blind17,blind18,blind21}. The second gourp is double-server BQC model \cite{blind2,blind11,blind20} and the third gourp is triple-server BQC model \cite{blind19}.
 In single-server BQC model, the client Alice is required to has the quantum ability of generating and distributing the single quantum states.
 In double-server BQC model and triple-server BQC model, the client Alice can be completely classical.

\begin{figure}[!h]
\begin{center}
\includegraphics[width=8cm,angle=0]{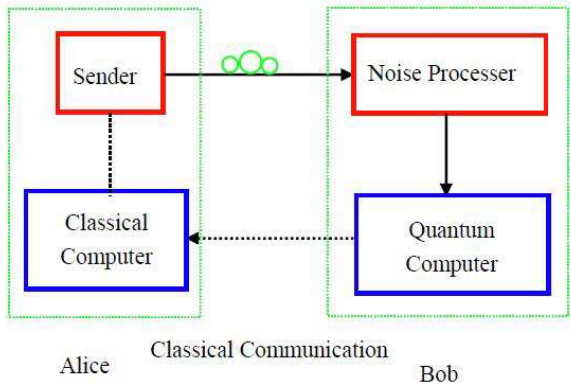}
\caption{Schematic of the anti-noise BQC protocol. After Alice prepares and encodes the single-photon state $|+_{\theta_{j}}\rangle$ to resist the collective noise using the Sender modular. Bob distills the polluted single-photon state using the Noise Processer modular, before starting the BQC protocol.}
\end{center}
\end{figure}

In  single-server BQC protocol, the client Alice should distribute the single-photon states to the server Bob. In previous single-server protocols, the quantum channel is assumed to be ideal.  However, the ideal quantum channel does not exist, and all the quantum states will suffer from noise. The environment noise will make the quantum state become error, and it will also  make the distributed photons loss.  In quantum communication, various error correction and error rejection
methods are proposed \cite{collective0,collective1,collective2,collective3,collective4,collective5}. For instance,  Walton \emph{et al.}
proposed a scheme for rejecting the errors introduced by
noise with decoherence-free subspaces \cite{collective0}. In 2005, Kalamidas proposed two interesting linear-optical single photon
schemes to reject and correct arbitrary qubit errors
without additional particles \cite{collective1}. By adding one extra photon with a fixed polarization, a  distribution scheme of polarization states of a single photon over a collective-noise
channel was proposed \cite{collective2}. In 2007, Li and Deng described a faithful qubit transmission scheme with linear optics against collective noise without ancillary qubits \cite{collective3}.
On the other hand, quantum state amplification is an efficient tools to resist the photon loss \cite{amplification1,amplification2,amplification3,amplification4,amplification5,amplification6,amplification7,amplification8}.
The quantum state amplification can be increase the probability of single photon and decrease the probability of photon loss.\\

\begin{figure}[!h]
\begin{center}
\includegraphics[width=6cm,angle=0]{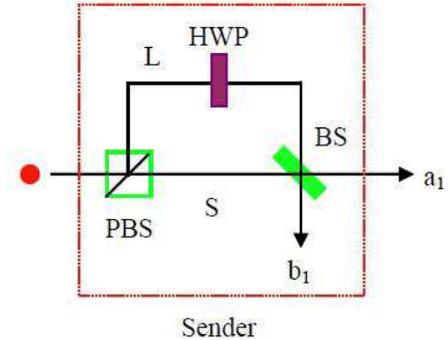}
\caption{ Schematic of the Sender modular as shown in Fig. 1. BS is the 50:50 beam splitter and PBS is the polarization beam splitter. HWP is the half-wave plate which can convert $|H\rangle$ polarization photon to $|V\rangle$ polarization photon and vice versa. $L$ and $S$ are the long and short arm of the photon.}
\end{center}
\end{figure}
Practical BQC protocol should also works under the noise environment. In double-server BQC,  Morimae and Fujii first described an efficient secure entanglement distillation for double-Server BQC \cite{blind11}. They showed that it is possible to perform entanglement distillation
in the double-server scheme without degrading the security of blind quantum computing. In 2015, we proposed the deterministic entanglement distillation for
secure double-server BQC \cite{blind20}.  In single-server BQC, recently, Takeuchi \emph{et al.} first considered the model of single-server BQC over a collective-noise channel, which is called DFS-BQC \cite{blind21}. They described three variations of DFS-BQC protocols, combined the ideas based on the DFS and the BFK protocol.
In this paper, we describe another anti-noise BQC protocol, based on the original BFK protocol \cite{blind2}.
This protocol has some advantages. First, Alice does not require to generate the Bell pair or coherent light, and only to distribute and operate the single photon with linear optics. which reduces the client's economic cost. Second, this protocol not only can protect the state from the collective noise, but also can distill the single photon from the photon loss.  Third, the noise setup in Bob is based on the feasible linear optics.

\section{Basic model of anti-noise BQC protocol}
Before we explain this protocol, we first briefly describe the original BFK protocol. It runs as follows \cite{blind2}: a) Client Alice first prepares $n$ rotated qubits $\{|+_{\theta_{j}}\rangle \equiv (|0\rangle+ e^{i\theta_{j}}|1\rangle)/\sqrt{2}\}^{n}_{j=1}$ and distributes to the server Bob.
Here $\theta_{j}\in \{k\pi/4 \mid k\in Z, 0\leq k\leq7\}$ and  $E$ is the set of edges of
$G$ and CZ$_{i;j}$ is the CZ gate between the $i$th and $j$th qubits. b) Bob prepares the Graph state $G$, which Alice tells her. Here $|G\{\theta_{j}\}\rangle\equiv(\bigotimes_{i,j}\in E)CZ_{i,j}$. c) Bob performs the measurement on the $j$th qubit according to measurement angle $\xi_{j}=\theta_{j}+\phi'_{j}+r_{j}\pi$, which Alice tells her. Here $r_{j}$ is a random number and $r_{j}\in\{0,1\}$.
 $\phi'_{}$ is the modified version of $\phi_{j}$ according to the previous measurement results. d) Bob sends the measurement results to Alice and Alice completes the computation with classical computer.

The basic model of this anti-noise BQC protocol is shown in Fig. 1. In the side of Alice, Alice first prepare $n$ rotated qubits $\{|+_{\theta_{j}}\rangle \equiv (|0\rangle+ e^{i\theta_{j}}|1\rangle)/\sqrt{2}\}^{n}_{j=1}$.
In an optical system, we denote the  horizontal polarization photon $|H\rangle$ as $|0\rangle$ and vertical polarization photon $|V\rangle$ as $|1\rangle$, respectively. In traditional BFK protocol, the single-photon state  $|+_{\theta_{j}}\rangle$ is sent to Bob directly.  In this protocol, Alice first encodes the state $|+_{\theta_{j}}\rangle$ as shown in Fig. 2.
\begin{eqnarray}
|+_{\theta_{j}}\rangle=\frac{1}{\sqrt{2}}(|H\rangle+ e^{i\theta_{j}}|V\rangle)\rightarrow \frac{1}{\sqrt{2}}(|H_{S}\rangle+ e^{i\theta_{j}}|H_{L}\rangle)\nonumber\\
\rightarrow \frac{1}{2} (|H_{S}\rangle_{a_{1}}+ e^{i\theta_{j}}|H_{L}\rangle_{a_{1}})+\frac{1}{2} (|H_{S}\rangle_{b_{1}}- e^{i\theta_{j}}|H_{L}\rangle_{b_{1}})
\end{eqnarray}
The photon will suffer from the noise, which will make
\begin{eqnarray}
|H\rangle_{a_{1}}\rightarrow \alpha |H\rangle_{a_{1}}+\beta |V\rangle_{a_{1}},
\end{eqnarray}
and
\begin{eqnarray}
|H\rangle_{b_{1}}\rightarrow \tau |H\rangle_{b_{1}}+\delta |V\rangle_{b_{1}}.
\end{eqnarray}
Here $|\alpha|^{2}+|\beta|^{2}=1$, and $|\tau|^{2}+|\delta|^{2}=1$. $a_{1}$ and $b_{1}$ are the spatial modes as shown in Fig. 2.
$S$ is the short arm and $L$ is the long arm.
The noise model is also the collective noise model \cite{blind21}.
Therefore, after transmission, if the photon does not loss, the state $|+_{\theta_{j}}\rangle$ becomes
\begin{eqnarray}
&&|+_{\theta_{j}}\rangle\rightarrow|+_{\theta_{j}}\rangle'=\frac{1}{2} [(\alpha |H_{S}\rangle_{a_{1}}+\beta |V_{S}\rangle_{a_{1}}\nonumber\\
&+& e^{i\theta_{j}}(\alpha |H_{L}\rangle_{a_{1}}+\beta |V_{L}\rangle_{a_{1}})]
+\frac{1}{2} [(\tau |H_{S}\rangle_{b_{1}}+\delta |V_{S}\rangle_{b_{1}}\nonumber\\
&-& e^{i\theta_{j}}(\tau |H_{L}\rangle_{b_{1}}+\delta |V_{L}\rangle_{b_{1}})].\label{single1}
\end{eqnarray}

Certainly, the single-photon state $|+_{\theta_{j}}\rangle$ may also suffer from the photon loss and become a vacuum state $|vac\rangle$. Generally,
Bob will receive a mixed state $\rho_{\theta_{j}}$, which can be written as
\begin{eqnarray}
\rho_{\theta_{j}}=F|+_{\theta_{j}}\rangle''\langle +_{\theta_{j}}|+(1-F)|vac\rangle\langle vac|.\label{mixed}
\end{eqnarray}

\begin{figure}[!h]
\begin{center}
\includegraphics[width=8cm,angle=0]{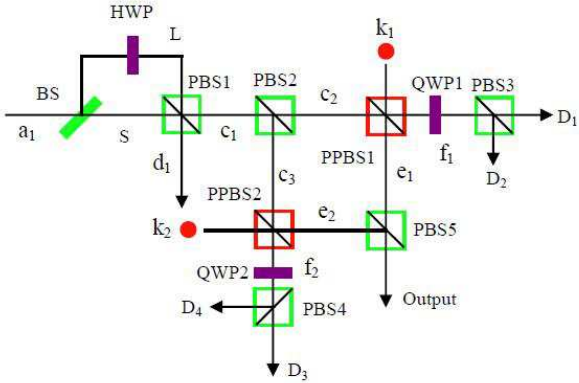}
\caption{Schematic of the Noise Processer as shown in Fig. 1. Bob requires  polarized Bell state in $k_{1}k_{2}$ spatial mode as auxiliary. QWP is the quarter wave plate which acts as the Hadamard operation. PPBS is the partial polarization beam splitter.}
\end{center}
\end{figure}
Here $F$ denotes the transmission efficiency of the photon. From Eq. (\ref{mixed}), before Bob starting the BQC protocol, he should distill the mixed state $\rho_{\theta_{j}}$ and obtain the original state $|+_{\theta_{j}}\rangle$ deterministically. The Noise Processer showed in Fig. 1 is to complete the task. The Noise Processor is detailed in Fig. 3.
From Eq. (\ref{single1}), if the single photon does not lose, it will be in the spatial modes $a_{1}$ or  $b_{1}$ with equal probability of 50\%.
The noise processor in Fig. 3 shows the distillation in spatial mode $a_{1}$. If the photon is in spatial mode $b_{1}$, Bob can distill it with the same setup.
We take the photon in $a_{1}$ spatial mode for example. As shown in Fig. 3, the quantum state can be evolved as
\begin{eqnarray}
&&(\alpha |H_{S}\rangle_{a_{1}}+\beta |V_{S}\rangle_{a_{1}})+ e^{i\theta_{j}}(\alpha |H_{L}\rangle_{a_{1}}+\beta |V_{L}\rangle_{a_{1}})\nonumber\\
&\rightarrow&\frac{1}{\sqrt{2}}(\alpha |H_{SS}\rangle+\alpha |V_{SL}\rangle
+\beta |H_{SL}\rangle+\beta |V_{SS}\rangle)\nonumber\\
&+&\frac{1}{\sqrt{2}}e^{i\theta_{j}}(\alpha |H_{LS}\rangle+\alpha |V_{LL}\rangle+\beta |H_{LL}\rangle+\beta |V_{LS}\rangle)\nonumber\\
&=&\frac{1}{\sqrt{2}}(\alpha |H_{SS}\rangle_{c_{1}}+\alpha |V_{SL}\rangle_{c_{1}}+\beta |H_{SL}\rangle_{d_{1}}+\beta |V_{SS}\rangle_{d_{1}})\nonumber\\
&+&\frac{1}{\sqrt{2}}e^{i\theta_{j}}(\alpha |H_{LS}\rangle_{c_{1}}+\alpha |V_{LL}\rangle_{c_{1}}+\beta |H_{LL}\rangle_{d_{1}}+\beta |V_{LS}\rangle_{d_{1}})\nonumber\\
&=&\frac{\alpha}{\sqrt{2}}(e^{i\theta_{j}}|H_{LS}\rangle_{c_{1}}+|V_{SL}\rangle_{c_{1}})\nonumber\\
&+&\frac{\beta}{\sqrt{2}}(|H_{SL}\rangle_{d_{1}}+e^{i\theta_{j}}|V_{LS}\rangle_{d_{1}})\nonumber\\
&+&\frac{1}{\sqrt{2}}(\alpha|H_{SS}\rangle_{c_{1}}+\beta |V_{SS}\rangle_{d_{1}})\nonumber\\
&+&\frac{1}{\sqrt{2}}e^{i\theta_{j}}(\alpha |V_{LL}\rangle_{c_{1}}+\beta |H_{LL}\rangle_{d_{1}}).\label{decode1}
\end{eqnarray}
From Eq. (\ref{decode1}), if the photon does not lose, it will be in the different arriving time, i. e., $SS$, $LS (SL)$, or $LL$, respectively.
Therefore, Bob can get the uncorrupted
states $|-_{\theta_{j}}\rangle$ in spatial mode $c_{1}$, or $|+_{\theta_{j}}\rangle$ in spatial mode $d_{1}$ in the determinate time corresponding to $SL$
and $LS$.
Here $|-_{\theta_{j}}\rangle=e^{i\theta_{j}}|H\rangle+|V\rangle$. One can perform a bit-flip operation $\sigma_{x}=|H\rangle\langle V|+|V\rangle\langle H|$ to convert $|-_{\theta_{j}}\rangle$ to $|+_{\theta_{j}}\rangle$.
From Eq. (\ref{single1}), if the single photon is in the spatial mode $b_{1}$, Bob can deal  with the same principle, using BS, HWP and PBS. Therefore, if the photon does not lose, combining with the single photon being in the spatial mode  $b_{1}$ with the same probability, Bob will obtain a single-photon entangled state in the time bin $SL (SL)$ as
\begin{eqnarray}
|\varphi\rangle&=&\frac{\alpha}{\sqrt{2}}|+_{\theta_{j}}\rangle_{c_{1}}|0\rangle_{d_{1}}|0\rangle_{m_{1}}|0\rangle_{n_{1}}\nonumber\\
&+&\frac{\beta}{\sqrt{2}}|0\rangle_{c_{1}}|+_{\theta_{j}}\rangle_{d_{1}}|0\rangle_{m_{1}}|0\rangle_{n_{1}}\nonumber\\
&+&\frac{\tau}{\sqrt{2}}|0\rangle_{c_{1}}|0\rangle_{d_{1}}|+_{\theta_{j}}\rangle_{m_{1}}|0\rangle_{n_{1}}\nonumber\\
&+&\frac{\delta}{\sqrt{2}}|0\rangle_{c_{1}}|0\rangle_{d_{1}}|0\rangle_{m_{1}}|+_{\theta_{j}}\rangle_{n_{1}}.\label{receive}
\end{eqnarray}
Here the spatial modes $m_{1}$ and $n_{1}$ are shown in Fig. 4. It means that if the photon is in the spatial mode $b_{1}$ as shown in Fig. 2, Bob can deal with the collective noise with the setup of Fig. 4.
\begin{figure}[!h]
\begin{center}
\includegraphics[width=7cm,angle=0]{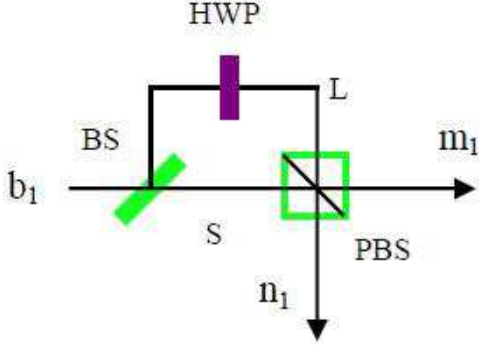}
\caption{Schematic of decoding if the single photon is in the spatial mode $b_{1}$.}
\end{center}
\end{figure}
Combined with the case of photon loss, by selecting the time bin $SL (SL)$,  state in Eq. (\ref{mixed}) can be rewritten as
\begin{eqnarray}
\rho'_{\theta_{j}}=F|\varphi\rangle\langle \varphi|+(1-F)|vac\rangle\langle vac|.\label{mixed1}
\end{eqnarray}
From Eq. (\ref{mixed1}), the next step of Bob is to distill $|\varphi\rangle$ from the mixed state deterministically. Here we exploit the linear noiseless
amplification (NLA) to complete the task. Here we  introduces a pair of ancillary polarized photons of the form
\begin{eqnarray}
|\phi_{1}\rangle=\frac{1}{\sqrt{2}}(|H\rangle_{k_{1}}|H\rangle_{k_{2}}+|V\rangle_{k_{1}}|V\rangle_{k_{2}}).
 \end{eqnarray}
As shown in Fig. 3, the partially polarization beam splitter (PPBS$_{1}$) can reflect the vertically polarized photon totally, while reflect the horizontally polarized photon with the coefficient of $\gamma$, and transmit it with the coefficient of $1-\gamma$. Another  PPBS$_{2}$ can transmit the horizontally polarized photon totally, while reflect the vertically polarized photon with the coefficient $\gamma$, and transmit it with the coefficient of $1-\gamma$.
   For instance, the PPBS$_{1}$ can make \cite{amplification5}
\begin{eqnarray}
\hat{a}^{\dag}_{c_{1},H}|0\rangle &\rightarrow& \gamma\hat{a}^{\dag}_{out,H}|0\rangle+\sqrt{1-\gamma^{2}}\hat{a}^{\dag}_{D1,H}|0\rangle,\nonumber\\
\hat{a}^{\dag}_{k_{1},H}|0\rangle &\rightarrow& -\gamma\hat{a}^{\dag}_{D1,H}|0\rangle+\sqrt{1-\gamma^{2}}\hat{a}^{\dag}_{out,H}|0\rangle,\nonumber\\
\hat{a}^{\dag}_{k_{1},V}|0\rangle&\rightarrow& -\hat{a}^{\dag}_{D1,V}|0\rangle.
\end{eqnarray}
Here $\hat{a}^{\dag}$ is the creation operator. The subscripts $c_{1}$, $out$, $k_{1}$ and $k_{2}$ are the spatial modes as shown in Fig. 3.
Therefore, by using PPBS$_{1}$ and PPBS$_{2}$, we can obtain the relationship
\begin{eqnarray}
|0_{c_{1}}H_{k_{1}}H_{k_{2}}\rangle &\rightarrow& \frac{\gamma}{2}|0_{out}\rangle(|H_{D_{1}}H_{D_{3}}\rangle+|H_{D_{1}}V_{D_{4}}\rangle\nonumber\\
&+&|V_{D_{2}}H_{D_{3}}\rangle+|V_{D_{2}}V_{D_{4}}\rangle),\nonumber\\
|0_{c_{1}}V_{k_{1}}V_{k_{2}}\rangle &\rightarrow& \frac{\gamma}{2}|0_{out}\rangle(|H_{D_{1}}H_{D_{3}}\rangle-|H_{D_{1}}V_{D_{4}}\rangle\nonumber\\
&-&|V_{D_{2}}H_{D_{3}}\rangle+|V_{D_{2}}V_{D_{4}}\rangle),\nonumber\\
|H_{c_{1}}H_{k_{1}}H_{k_{2}}\rangle &\rightarrow& \frac{(2\gamma^{2}-1)}{2}|H_{out}\rangle(|H_{D_{1}}H_{D_{3}}\rangle+|H_{D_{1}}V_{D_{4}}\rangle\nonumber\\
&+&|V_{D_{2}}H_{D_{3}}\rangle+|V_{D2}V_{D4}\rangle),\nonumber\\
|H_{c_{1}}V_{k_{1}}V_{k_{2}}\rangle &\rightarrow& \frac{\gamma^{2}}{2}|H_{out}\rangle(|H_{D_{1}}H_{D_{3}}\rangle-|H_{D_{1}}V_{D_{4}}\rangle\nonumber\\
&-&|V_{D_{2}}H_{D_{3}}\rangle+|V_{D_{2}}V_{D_{4}}\rangle),\nonumber\\
|V_{c_{1}}H_{k_{1}}H_{k_{2}}\rangle &\rightarrow& \frac{\gamma^{2}}{2}|V_{out}\rangle (|H_{D_{1}}H_{D_{3}}\rangle+|H_{D_{1}}V_{D_{4}}\rangle\nonumber\\
&+&|V_{D_{2}}H_{D_{3}}\rangle+|V_{D_{2}}V_{D_{4}}\rangle),\nonumber\\
|V_{c_{1}}V_{k_{1}}V_{k_{2}}\rangle &\rightarrow& \frac{(2\gamma^{2}-1)}{2}|V_{out}\rangle(|H_{D_{1}}H_{D_{3}}\rangle-|H_{D_{1}}V_{D_{4}}\rangle\nonumber\\
&-&|V_{D_{2}}H_{D_{3}}\rangle+|V_{D_{2}}V_{D_{4}}\rangle).\nonumber\\
\end{eqnarray}
 The mixed state $\rho'_{\theta_{j}}$ combined with the polarization Bell state $|\phi_{1}\rangle$ can be described as follows. With the probability of $F$, it is in the state $|\varphi\rangle\otimes|\phi_{1}\rangle$ and with the probability of $1-F$, it is in the state $|vac\rangle\otimes|\phi_{1}\rangle$.
We first discuss the item $|\varphi\rangle\otimes|\phi_{1}\rangle$. It evolves as
\begin{eqnarray}
&&|\varphi\rangle\otimes|\phi_{1}\rangle=[\frac{\alpha}{\sqrt{2}}|+_{\theta_{j}}\rangle_{c_{1}}|0\rangle_{d_{1}}|0\rangle_{m_{1}}|0\rangle_{n_{1}}\nonumber\\
&+&\frac{\beta}{\sqrt{2}}|0\rangle_{c_{1}}|+_{\theta_{j}}\rangle_{d_{1}}|0\rangle_{m_{1}}|0\rangle_{n_{1}}\nonumber\\
&+&\frac{\tau}{\sqrt{2}}|0\rangle_{c_{1}}|0\rangle_{d_{1}}|+_{\theta_{j}}\rangle_{m_{1}}|0\rangle_{n_{1}}\nonumber\\
&+&\frac{\delta}{\sqrt{2}}|0\rangle_{c_{1}}|0\rangle_{d_{1}}|0\rangle_{m_{1}}|+_{\theta_{j}}\rangle_{n_{1}}]\nonumber\\
&\otimes&\frac{1}{\sqrt{2}}(|H\rangle_{k_{1}}|H\rangle_{_{2}}+|V\rangle_{k_{1}}|V\rangle_{k_{2}}).\label{collapse}
\end{eqnarray}
The first item  evolves as
\begin{eqnarray}
&&\frac{\alpha}{\sqrt{2}}|+_{\theta_{j}}\rangle_{c_{1}}|0\rangle_{d_{1}}|0\rangle_{m_{1}}|0\rangle_{n_{1}}\nonumber\\
&\otimes&\frac{1}{\sqrt{2}}(|H\rangle_{k_{1}}|H\rangle_{_{2}}+|V\rangle_{k_{1}}|V\rangle_{k_{2}})\nonumber\\
&=& \frac{\alpha}{2\sqrt{2}}(|H\rangle_{c_{1}}|H\rangle_{k_{1}}|H\rangle_{k_{2}}+|H\rangle_{c_{1}}|V\rangle_{k_{1}}|V\rangle_{k_{2}}\nonumber\\
&+&e^{i\theta}|V\rangle_{c_{1}}|H\rangle_{k_{1}}|H\rangle_{k_{2}}+e^{i\theta}|V\rangle_{c_{1}}|V\rangle_{k_{1}}|V\rangle_{k_{2}})\nonumber\\
&\otimes&|0\rangle_{d_{1}}|0\rangle_{m_{1}}|0\rangle_{n_{1}}\nonumber\\
&\rightarrow&\frac{\alpha}{2\sqrt{2}}[\frac{(2\gamma^{2}-1)}{2}|H_{out}\rangle(|H_{D_{1}}H_{D_{3}}\rangle+|H_{D_{1}}V_{D_{4}}\rangle\nonumber\\
&+&|V_{D_{2}}H_{D_{3}}\rangle+|V_{D2}V_{D4}\rangle)|0\rangle_{d_{1}}|0\rangle_{m_{1}}|0\rangle_{n_{1}}\nonumber\\
&+&\frac{\gamma^{2}}{2}|V_{out}\rangle (|H_{D_{1}}H_{D_{3}}\rangle+|H_{D_{1}}V_{D_{4}}\rangle\nonumber\\
&+&|V_{D_{2}}H_{D_{3}}\rangle+|V_{D_{2}}V_{D_{4}}\rangle)|0\rangle_{d_{1}}|0\rangle_{m_{1}}|0\rangle_{n_{1}}\nonumber\\
&+&e^{i\theta}\frac{\gamma^{2}}{2}|V_{out}\rangle (|H_{D_{1}}H_{D_{3}}\rangle+|H_{D_{1}}V_{D_{4}}\rangle\nonumber\\
&+&|V_{D_{2}}H_{D_{3}}\rangle+|V_{D_{2}}V_{D_{4}}\rangle)|0\rangle_{d_{1}}|0\rangle_{m_{1}}|0\rangle_{n_{1}}\nonumber\\
&+&\frac{(2\gamma^{2}-1)}{2}|V_{out}\rangle(|H_{D_{1}}H_{D_{3}}\rangle-|H_{D_{1}}V_{D_{4}}\rangle\nonumber\\
&-&e^{i\theta}|V_{D_{2}}H_{D_{3}}\rangle+|V_{D_{2}}V_{D_{4}}\rangle)|0\rangle_{d_{1}}|0\rangle_{m_{1}}|0\rangle_{n_{1}}].\label{collapse1}
\end{eqnarray}
The second item can evolve as
\begin{eqnarray}
&&\frac{\beta}{\sqrt{2}}|0\rangle_{c_{1}}|+_{\theta_{j}}\rangle_{d_{1}}|0\rangle_{m_{1}}|0\rangle_{n_{1}}\nonumber\\
&\otimes&\frac{1}{\sqrt{2}}(|H\rangle_{k_{1}}|H\rangle_{_{2}}+|V\rangle_{k_{1}}|V\rangle_{k_{2}})\nonumber\\
&\rightarrow&\frac{\beta\gamma}{4}(|0_{out}\rangle(|H_{D_{1}}H_{D_{3}}\rangle+|H_{D_{1}}V_{D_{4}}\rangle\nonumber\\
&+&|V_{D_{2}}H_{D_{3}}\rangle+|V_{D_{2}}V_{D_{4}}\rangle)\nonumber\\
&+&|0_{out}\rangle(|H_{D_{1}}H_{D_{3}}\rangle-|H_{D_{1}}V_{D_{4}}\rangle+|V_{D_{2}}H_{D_{3}}\rangle+|V_{D_{2}}V_{D_{4}}\rangle)\nonumber\\
&\otimes&|+_{\theta_{j}}\rangle_{d_{1}}|0\rangle_{m_{1}}|0\rangle_{n_{1}}\nonumber\\
&=&\frac{\beta\gamma}{4}|0_{out}\rangle|+_{\theta_{j}}\rangle_{d_{1}}|0\rangle_{m_{1}}|0\rangle_{n_{1}}(|H_{D_{1}}H_{D_{3}}\rangle+|V_{D_{2}}V_{D_{4}}\rangle).\nonumber\\\label{collapse2}
\end{eqnarray}
The third item can evolve as
\begin{eqnarray}
&&\frac{\tau}{\sqrt{2}}|0\rangle_{c_{1}}|0\rangle_{d_{1}}|+_{\theta_{j}}\rangle_{m_{1}}|0\rangle_{n_{1}}\nonumber\\
&\otimes&\frac{1}{\sqrt{2}}(|H\rangle_{k_{1}}|H\rangle_{_{2}}+|V\rangle_{k_{1}}|V\rangle_{k_{2}})\nonumber\\
&\rightarrow&\frac{\tau\gamma}{4}|0\rangle_{c_{1}}|0\rangle_{d_{1}}|+_{\theta_{j}}\rangle_{m_{1}}|0\rangle_{n_{1}}(|H_{D_{1}}H_{D_{3}}\rangle+|V_{D_{2}}V_{D_{4}}\rangle).\nonumber\\\label{collapse3}
\end{eqnarray}
The forth item can evolve as
\begin{eqnarray}
&&\frac{\delta}{\sqrt{2}}|0\rangle_{c_{1}}|0\rangle_{d_{1}}|0\rangle_{m_{1}}|+_{\theta_{j}}\rangle_{n_{1}}\nonumber\\
&\otimes&\frac{1}{\sqrt{2}}(|H\rangle_{k_{1}}|H\rangle_{_{2}}+|V\rangle_{k_{1}}|V\rangle_{k_{2}})\nonumber\\
&\rightarrow&\frac{\delta\gamma}{4}|0\rangle_{c_{1}}|0\rangle_{d_{1}}|0\rangle_{m_{1}}|+_{\theta_{j}}\rangle_{n_{1}}(|H_{D_{1}}H_{D_{3}}\rangle+|V_{D_{2}}V_{D_{4}}\rangle).\nonumber\\\label{collapse4}
\end{eqnarray}
 Interestingly, from Eq. (\ref{collapse1}) to (\ref{collapse4}), if they pick up the case that the single-photon detectors D$_{1}$D$_{4}$ or D$_{2}$D$_{3}$ register one photon respectively, they will obtain the state $|+\rangle_{\theta_{j}}$ deterministically in the output mode, for the cases in Eq. (\ref{collapse2}) to (\ref{collapse4}) can not lead both single-photon detectors D$_{1}$D$_{4}$ or D$_{2}$D$_{3}$ register one photon, and only the item in Eq. (\ref{collapse1}) can satisfy the selection condition.

Once Bob obtain the single-photon state $|+\rangle_{\theta_{j}}$ deterministically, he can start to perform the BQC protocol.  In this way, the whole BQC protocol can be modified as follows: 1) Alice prepares $n$ rotated qubits $\{|+\rangle_{\theta_{j}}\}^{n}_{j=1}$. 2) Alice encodes the photon $|+\rangle_{\theta_{j}}$ with linear optics, as shown in Fig. 2. 3) Alice distributes the photon to Bob, which will suffer from collective noise and photon loss. 4) Bob distill the polluted single-photon states with Noise Processer setup, as shown in Fig. 3. 5) Bob prepares the Graph state $G$. 6) Bob performs the measurement on the $j$th qubit. 7) Bob sends the measurement results to Alice and Alice  completes the computation with classical computer.

\section{Discussion and conclusion}
So far, we have completely explained the anti-noise BQC protocol. In original BQC protocol, Alice prepares and distributes the state $|+\rangle_{\theta_{j}}$ directly. Bob also receives the $|+\rangle_{\theta_{j}}$ and perform the BQC subsequently, for they do not consider the noise environment. Similar to the pioneer work of Ref.\cite{blind21} in collective-noise BQC protocol, Alice and Bob should perform the pretreatment for noise, before starting the BQC protocol, following the approaches suggested in Refs.\cite{collective3,amplification5}. In BQC protocol, two essential properties are correctness and blindness. The correctness means that the output of the protocol in Alice's desired one as long as Alice and Bob follow the procedure of the protocol is faithfully. On the other hand, the blindness means that Bob cannot know any information about Alice's inputs, algorithm, and outputs, whenever Alice follows the procedure of the protocol. Obviously, this protocol is correctness for Bob can obtain the faithful qubits after the Noise Processer. On the other hand, this protocol is also blindness. The information sent from Alice to Bob is $|+\rangle_{\theta_{j}}$, which is decided by $\theta_{j}$.  Bob does not know the exact information of $\theta_{j}$, which ensures that the protocol is blindness.

We can calculate the total success probability of this protocol. From Fig. 2, we explain this protocol by selecting the case that the photon being in the spatial mode $a_{1}$ with the probability of 50\%. Actually, if the photon is in the spatial mode $b_{1}$, they can perform the protocol with the same principle. That is to say, if the photon does not lose and only suffer from the collective noise, by picking up the suitable arriving time $SL (LS)$ as shown in Fig. 4, the total success probability is 50\%. On the other hand, as shown in Fig. 3, such setup essentially  distill the photons in the spatial modes $c_{1}$, i. e., the first item in Eq. (\ref{receive}). Actually, in Eq. (\ref{receive}),  other three items which  contains the single photon that can also be verified by adding three auxiliary polarized Bell states.   The final success probability of this protocol can be calculated as
\begin{eqnarray}
P=\frac{F(5\gamma^{4}-4\gamma^{2}+1)}{32}.
\end{eqnarray}

In the previous work of Ref.\cite{blind21}, they presented three important BQC protocols over a collective-noise channel. The first protocol is entanglement-based protocol. The second is single-photon-based  protocol and the third is the coherent-light-assisted protocol. The common characteristics of three protocol is that they require auxiliary resources, such as entanglement, single photon or coherent light. Moreover, these protocols suppose that Bob has powerful QND measurement, which can distinguish the arriving photons deterministically. However, they do not explain how to realize such QND measurement.
In current technology, QND measurement, such as exploiting cross-Kerr nonlinearity has  widely discussed in quantum information processing \cite{kerr1,kerr2,kerr3,kerr4,kerr5}, but it is still a big challenge in experiment \cite{kerr6} and it will greatly increase the computation cost.

In conclusion, we have described an efficient anti-noise BQC protocol.  Different from previous work, this protocol has several advantages. First, this protocol does not require any auxiliary resources, which makes the client is economic. Second, this protocol not only can protect the state from the collective noise, but also from the photon loss.  Third, the Noise Processer for Bob is based on the linear optics, and it is also feasible in experiment.

\section*{ACKNOWLEDGEMENTS}
This work was supported by the National Natural Science Foundation
of China under Grant  Nos. 11474168 and 61401222, the Qing Lan Project in Jiangsu Province, and a Project
Funded by the Priority Academic Program Development of Jiangsu
Higher Education Institutions.\\


\begin{thebibliography}{99}
\bibitem{shor}  P. W. Shor, Foundations of Computer Science, 1994 Proceedings., 35th Annual Symposium on. IEEE,  124-134 (1994).

\bibitem{grover} L. K. Grover,  Proceedings of the twenty-eighth annual ACM symposium on Theory of computing. ACM, 212-219 (1996).

\bibitem{long} G. L. Long, Phys. Rev. A  \textbf{64}, 022307 (2001).

\bibitem{ion}J. I. Cirac and P. Zoller, Phys. Rev. Lett. \textbf{74}, 4091 (1995).

\bibitem{superconduct}Y. Makhlin, G. Sch\"{o}n, and A. Shnirman, Rev. Mod. Phys. \textbf{73},
357 (2001).

\bibitem{photon}C.-Y. Lu, \emph{et al.}, Nat. Phys. \textbf{3}, 91 (2007).

\bibitem{other1}J. Berezovsky, M. H. Mikkelsen, N. G. Stoltz, L. A. Coldren, and
D. D. Awschalom, Science \textbf{320}, 349 (2008); R. Hanson and D. D. Awschalom, Nature \textbf{453}, 1043 (2008).

\bibitem{blind1}A. Childs, Quantum Inf. Comput. \textbf{5}, 456 (2005).

\bibitem{blind2}A. Broadbent, J. Fitzsimons, and E. Kashefi, Proceedings of
the 50th Annual IEEE Symposium on Foundations of Computer
Science (IEEE, Piscataway, NJ, 2009), p. 517.

\bibitem{blind3}T. Morimae, V. Dunjko, and E. Kashefi, arXiv:1009.3486.

\bibitem{blind4}T. Morimae and K. Fujii, Nat. Commun. \textbf{3}, 1036 (2012).

\bibitem{blind5}J. Fitzsimons, and E. Kashefi, arXiv:1203.5217.

\bibitem{blind6}T. Morimae, Phys. Rev. Lett. \textbf{109}, 230502 (2012).

\bibitem{blind7}V. Dunjko, E. Kashefi, and A. Leverrier, Phys. Rev. Lett. \textbf{108},
200502 (2012).

\bibitem{blind8}T. Morimae and K. Fujii, Phys. Rev. A \textbf{87}, 050301(R) (2013).

\bibitem{blind9}T. Sueki, T. Koshiba, and T. Morimae, Phys. Rev. A \textbf{87}, 060301(R) (2013).

\bibitem{blind10}S. Barz, E. Kashefi, A. Broadbent, J. F. Fitzsimons, A. Zeilinger, and P. Walther, Science \textbf{335}, 303 (2012).

\bibitem{blind11}T. Morimae and K. Fujii, Phys. Rev. Lett. \textbf{111}, 020502 (2013).

\bibitem{blind12}V. Giovannetti, L. Maccone, T. Morimae, and T. G. Rudolph,
Phys. Rev. Lett. 111, 230501 (2013).

\bibitem{blind13}A. Mantri, C. A. P. Delgado, and J. F. Fitzsimons,
Phys. Rev. Lett. \textbf{111}, 230502 (2013).

\bibitem{blind14}T. Sueki, T. Koshiba, and T. Morimae, Phys. Rev. A \textbf{87},
060301(R) (2013).

\bibitem{blind15}T. Morimae and K. Fujii, Phys. Rev. A \textbf{87}, 050301(R) (2013).

\bibitem{blind16}K. A. G. Fisher, A. Broadbent, L. K. Shalm, Z. Yan, J. Lavoie,
R. Prevedel, T. Jennewein, and K. J. Resch, Nat. Commun. \textbf{5},
3074 (2014).

\bibitem{blind17}T. Morimae, Phys. Rev. A \textbf{89}, 060302(R) (2014).

\bibitem{blind18}A. Gheorghiu, E. Kashefi, and P. Wallden, New J. Phys. \textbf{17},
083040 (2015).
\bibitem{blind19} Q. Li,  W. H. Chan, C. Wu,  and  Z. Wen,  Phys. Rev. A \textbf{89}, 040302(R) (2014).

\bibitem{blind20}Y. B. Sheng, and L. Zhou, Sci. Rep. \textbf{15}, 7815 (2015).

\bibitem{blind21}V. Takeuchi, K. Fujii, R. Ikuta, T. Yamamoto, and N. Imoto, Phys. Rev. A \textbf{93}, 052307 (2016).

\bibitem{collective0} D. Walton, A. F. Abouraddy, A. V. Sergienko, B. E. A. Saleh, and M. C.
Teich, Phys. Rev. Lett. \textbf{91}, 087901 (2003).

\bibitem{collective1}D. Kalamidas, Phys. Lett. A \textbf{343}, 331(2005).

\bibitem{collective2}T. Yamamoto, J. Shimamura, S. K. \"{O}zdemir, M. Koashi, and N. Imoto,
Phys. Rev. Lett. \textbf{95}, 040503 (2005).

\bibitem{collective3}X. H. Li, F. G. Deng, and H. Y. Zhou, Appl. Phys. Lett. \textbf{91}, 144101 (2007).

\bibitem{collective4}Y. B. Sheng, and F. G. Deng, Phys. Rev. A \textbf{81}, 042332 (2010).

\bibitem{collective5}H. Kumagai, . Yamamoto, M. Koashi, and N. Imoto, Phys. Rev. A \textbf{87}, 052325 (2013).

\bibitem{amplification1}N. Gisin, S. Pironio, and N. Sangouard, Phys. Rev. Lett. \textbf{105},
070501 (2010).

\bibitem{amplification2}G. Y. Xiang, T. C. Ralph, A. P. Lund, N. Walk, and G. J. Pryde,
Nat. Photon. \textbf{4}, 316 (2010).

\bibitem{amplification3} C. I. Osorio,  N. Bruno,  N. Sangouard,  H. Zbinden,  N. Gisin, and R.T. Thew, Phys. Rev. A \textbf{86}, 023815 (2012).

\bibitem{amplification4}S. L. Zhang,  S. Yang, X.B. Zou,  B.S. Shi, and G.C. Guo,  Phys. Rev. A \textbf{86}, 034302 (2012).

\bibitem{amplification5} E., Meyer-Scott,  M., Bula,  K. Bartkiewicz,  A. \v{C}ernoch,  J. Soubusta, T. Jennewein, and K. Lemr,  Phys. Rev. A \textbf{88}, 012327 (2013).

\bibitem{amplification6}Y. B.  Sheng,  Y. Ou-Yang, L. Zhou, and L. Wang,  Quantum Inf. Process. \textbf{13}, 1595 (2014).

\bibitem{amplification7}Z. F. Feng,  Y.  Ou-Yang, L. Zhou, and Y.B. Sheng,
Opt. Commun. \textbf{340}, 80 (2015).

\bibitem{amplification8}N. Bruno,  V. Pini,   A. Martin,  V. B. Verma, S. W. Nam,  R. Mirin,  A. Lita,  F. Marsili,  B. Krozh, F. Bussi\`{e}res,
N. Sangouard,  H. Zbinden, N.  Gisin, and R. Thew,  Opt. Express \textbf{24}, 125 (2016).

\bibitem{kerr1} K. Nemoto and W. J. Munro,  Phys. Rev. Lett. \textbf{93}, 250502 (2004).

\bibitem{kerr2} B. He, Y. Ren, and J. A. Bergou, Phys. Rev. A  \textbf{79}, 052323 (2009)

\bibitem{kerr3}Y. B.  Sheng, and  L. Zhou,   Sci. Rep. \textbf{5}, 13453 (2015).

\bibitem{kerr4}D. Ding, F. L.  Yan,  and T. Gao, Sci. China-Phys. Mecha. \& Astro. \textbf{57}, 2098 (2014).

\bibitem{kerr5}L. Dong , J. X. Wang, Q. Y. Li, H. Z. Shen, H. K. Dong, X. M. Xiu, Y. J. Gao,  and H. O. Choo,  Phys. Rev. A \textbf{93}, 012308 (2016).

\bibitem{kerr6} P. Kok, W. J. Munro, K. Nemoto, T. C. Palph, J. P. Dowling, and G. J. Milburn,  Rev. Mod. Phys. \textbf{79}, 135 (2007).
\end{thebibliography}
\end{document}